\documentclass[rnote]{aa} 
\usepackage{graphicx}
\usepackage{txfonts}
\begin{document}
   \title{Upgrading electron temperature and electron density diagnostic diagrams of forbidden line emission}

   \author{B. Proxauf\inst{\ref{inst1}}\and S. {\"O}ttl\inst{\ref{inst1}}\and S. Kimeswenger\inst{\ref{inst2},\ref{inst1}}}

   \institute{Institute for Astro- and Particle Physics, Leopold Franzens Universit{\"{a}}t Innsbruck,
              Technikerstrasse 25, 6020 Innsbruck, Austria\newline
              \email{Bastian.Proxauf@student.uibk.ac.at, Silvia.Oettl@uibk.ac.at}\label{inst1}
   \and
   Insituto de Astronom{\'{i}}a, Universidad Cat{\'{o}}lica del Norte,
                Avenida Angamos 0610, Antofagasta, Chile\newline
              \email{Stefan.Kimeswenger@gmail.com}\label{inst2}
             }

   \date{Received 2$^{\rm nd}$ September, 2013; accepted 20$^{\rm th}$ November, 2013}

\sloppy
\abstract{
Diagnostic diagrams of forbidden lines have been a useful tool for observers in astrophysics for many decades now. They are used to obtain information on the basic physical properties of thin gaseous nebulae. Some diagnostic diagrams are in wavelength domains which were difficult to take either due to missing wavelength coverage or low resolution of older spectrographs. Furthermore, most of the diagrams were calculated using just the species involved as a single atom gas, although several are affected by well-known fluorescence mechanisms as well. Additionally the atomic data have improved up to the present time.
}
{Aim of this work was a recalculation of well-known, but also of sparsely used, unnoted diagnostics diagrams. The new diagrams provide observers with modern, easy-to-use recipes to determine electron temperature and densities.
}
{The new diagnostic diagrams are calculated using large grids of parameter space in the photoionization code CLOUDY. For a given basic parameter (e.g. electron density or temperature) the solutions with cooling-heating-equilibrium are chosen to derive the diagnostic diagrams. Empirical numerical functions are fitted to provide formulas usable in e.g. data reduction pipelines.
 }
{The resulting diagrams differ significantly from those used up to now and will improve the thermodynamic calculations. To our knowledge, for the first time detailed directly applicable fit formulas are given, leading to electron temperature or density from the line ratios.}
   {}


   \maketitle
%

\section{Introduction}

Diagnostic diagrams of forbidden line emission from astrophysical plasmas are widely accepted as a tool to derive electron temperature $T_{\rm e}$ and the (number) density $n_{\rm e}$. The physics is well understood (see Osterbrock \& Ferland \cite{O_F_06}, hereinafter OF06). But the hitherto published diagrams mostly do not take into account disturbance of the basic processes by other atoms - namely by fluorescence.

\medskip
\noindent{\it The electron temperature $T_{\rm e}$:\newline}
\indent To derive the electron temperature $T_{\rm e}$, at least two metastable levels above the ground level with different energies are required. Due to the change of the populations with $T_{\rm e}$, the line strengths are a direct indicator for its value. In contrast to the determination of the electron densities described below, the collisional de-excitation must not play a role here. This  boundary condition is often overlooked by observers when using the classical diagnostic diagrams in dense environments as well.
Typical representatives are [\ion{O}{iii}] $\lambda\lambda$\ (4958\AA~+ 5007\AA) / $\lambda$ 4363\AA\ using the exponential approximation in OF06 and  [\ion{Ar}{iii}] $\lambda\lambda$\ (7135\AA~+ 7751\AA) / $\lambda$ 5192\AA, published in Keenan et al. (\cite{Keenan_88}). As shown by Izotov et al. (\cite{izotov06}) and Nicholls et al. (\cite{nichols12}), especially the widely used exponential law of [\ion{O}{iii}] suffers from the approximations of temperature independent collision strengths.
\medskip

\noindent{\it The electron density $n_{\rm e}$:\newline}
\indent To derive the electron density $n_{\rm e}$, a pair of lines with nearly the same excitation state is chosen. Both of them have to be metastable. In case of thin media, without collisions of the ions with electrons, the line ratio is given simply by their Einstein coefficients. When density increases, the lines with longer lifetimes are affected by collisional de-excitation first. Towards the high-density limit, the ratios of the lifetimes give the line ratio. Due to the transition, the line ratio is an indicator for the density between these limits. The major representatives for these diagrams are [\ion{S}{ii}] $\lambda\lambda$ 6716\AA~/ 6732\AA\  in the calibration of OF06 and [\ion{O}{ii}] $\lambda\lambda$ 3726\AA~/ 3729\AA\ by Pradhan et al. (\cite{Pradhan06}). The latter pair is difficult to observe due to the small distance between the lines. Both ions have ionization energies (from the lower state) close to that of hydrogen. The line pair of [\ion{Ar}{iv}] $\lambda\lambda$ 4711\AA~/ 4740\AA, calibrated by Stanghellini \& Kaler (\cite{ArIV}, hereinafter SK89), is sparsely used. The blue line of the pair is very near to a \ion{He}{i} emission. On the other hand this ionization state fits well to the often used temperature determination using [\ion{O}{iii}] (see above).
\medskip

\noindent The models used here assume thermalization of the electrons. The thermalization of the electrons is a major additional boundary condition in our case. Towards extremely thin environments, as it can be found in the intergalactic or intracluster medium, these or similar diagnostic diagrams cannot be applied anymore. The influence of non-Maxwellian $\kappa$ distributions on solar wind plasmas was studied recently by Nicholls et al. (\cite{kappa}). These are plasmas with long-range pumping of electron energies. The timescales are similar to the collisional relaxation time (Leubner \cite{leubner}; Livadiotis \& McComas \cite{kappa09}) usually denoted to magnetic fields.
\medskip

We compiled in this work new diagrams for the two most frequently used diagrams and for the two rarely used diagrams based on argon. We used the  CLOUDY C13 code and the included modern atomic data (Ferland et al. \cite{Cloudy_13}). CLOUDY was chosen as it includes to our knowledge the most sophisticated physics for photoionized astrophysical plasmas. Further, we give empirical fit formulas for direct use for observers and for implementation into data reduction recipes.


\section{Setup and Calculations}
The three latest versions of the photoionization code CLOUDY, namely C08, C10 and C13, are used (Ferland et al. \cite{Cloudy_90}, \cite{Cloudy_13}) for our calculations. CLOUDY gives a very sophisticated framework, containing a large set of physical interactions in plasmas with mixed chemistry. Even dust particles are included. The latter are important for charge transfer (van Hoof et al. \cite{vanHoof00}) and allow molecular \element{H}$_2$ and neutral \ion{H}{i} gas to survive in the ionized regions of planetary nebulae (PNe) (Aleman \& Gruenwald \cite{AlGr04}). Including all those processes is essential for the usage of the diagrams by observers.
\medskip

In this setup, the temperature or the density (or both of them) are fixed in a thin gas slab, illuminated by a hot OB star at a distance of 0.1\,pc. In case of simple calculations for diagnostic diagrams of individual isolated ions in an electron gas, this setup with a pair of fixed values would be sufficient. However, a full physical description with fluorescence and forbidden line cooling requires a complete equilibrium of heating and cooling. Only then it is fully applicable to analyze observations of nebulae. Therefore the stellar temperature or the luminosity are varied until this equilibrium is found. We find no major variations, except for [\ion{O}{iii}] we find differences worth to be included properly. We attribute this to the effects of the Bowen fluorescence. The whole parameter space is scanned using the {\verb"vary"} function in C13. Therefore up to about 100 models are calculated for every data point along the curve. A dedicated C program is written as wrapper around the whole setup for the selection of the appropriate results. An example set of input files is described in the appendix of Proxauf (\cite{BSc}) and can be obtained electronically from the authors. At plasma temperatures above 26\,000\,K, no physically reasonable equilibrium for thermal dominated radiation processes by photoionization is possible. Forbidden line cooling dominates here. Thus we conclude that these temperatures cannot be reached in normal photoionized regions without other processes (e.g. shocks). Therefore our investigations end at slightly lower temperatures than those shown in literature before.

\section{Results}
\subsection{The Electron Temperature}

\begin{figure}
   \centering
   \includegraphics[width=8.8cm]{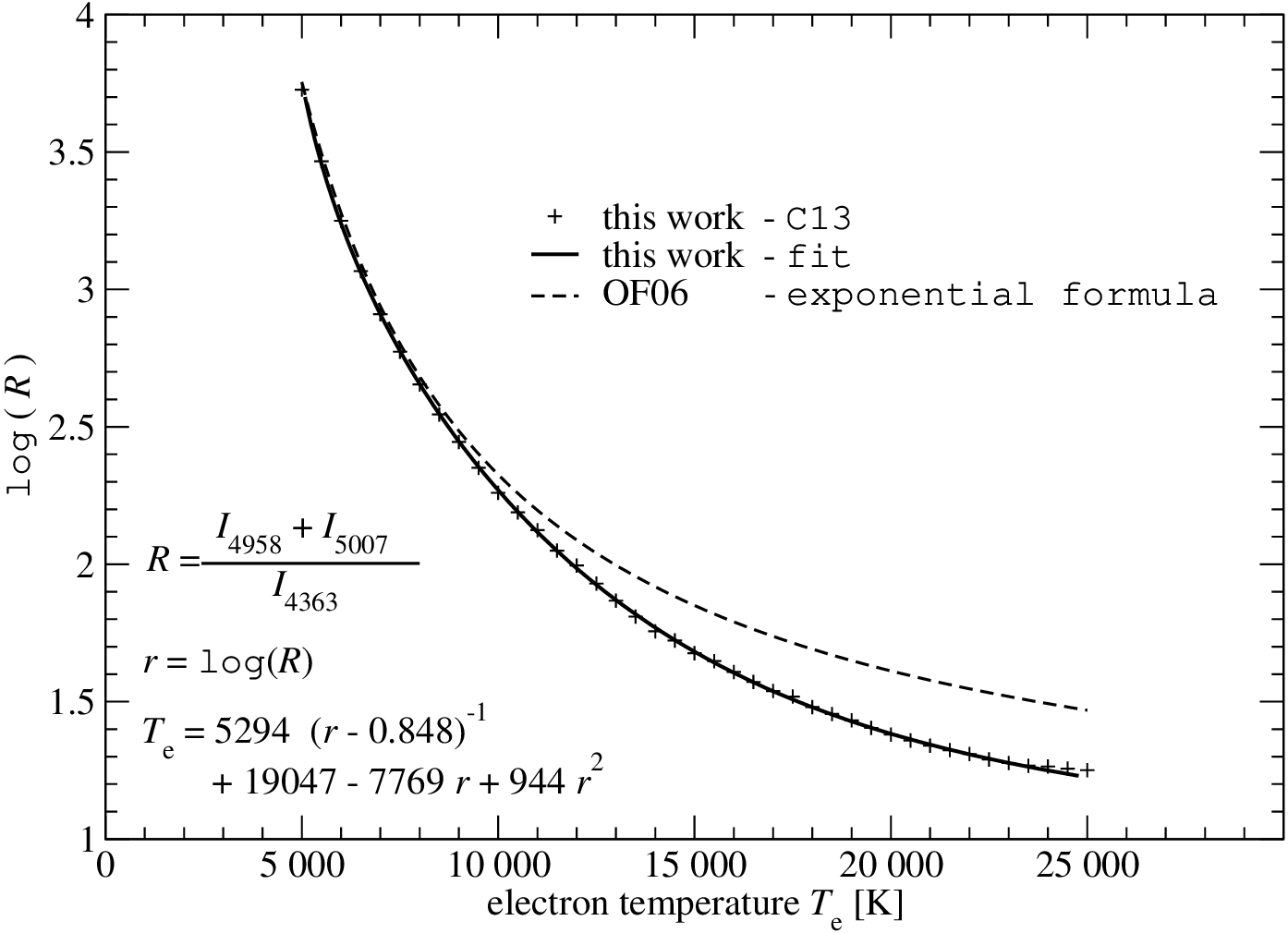}
      \caption{
   The diagnostic diagram for the electron temperature using the [\ion{O}{iii}] lines. Starting already at $T_{\rm e} \approx 10^4\,{\rm K}$, strong       deviations from the widely used exponential formula are evident.}
         \label{oiii_temperature.fig}
\end{figure}

To derive electron temperatures the line set [\ion{O}{iii}] $\lambda\lambda$ (4958\AA~+ 5007\AA) / $\lambda$ 4363\AA~from the two lower transitions $^1D_{2}\rightarrow\phantom{l}^3P_{2}$ (2.513$\rightarrow$0.038\,eV), $^1D_{2}\rightarrow\phantom{l}^3P_{1}$ (2.513$\rightarrow$0.014\,eV) and the upper transition $^1S_{0}\rightarrow\phantom{l}^1D_{2}$ (5.354$\rightarrow$2.513\,eV) is mostly used. This method is applicable up to electron densities of a few $10^4\,{\rm cm}^{-3}$. The exponential formula in OF06 has been derived using just the basic thermodynamic system of these lines. Pumping via the Bowen fluorescence causes additional populations of the upper level at higher temperatures. Moreover new atomic data is available. Using the atomic data of Palay et al. (\cite{palay12}), already Nicholls et al. (\cite{kappa}) point out that the exponential formula overestimates systematically the temperature. In Equ.~\ref{te_oiii} our conversion into temperature for the observed line ratio $R$ is given. Vice versa, as the observers have line ratios $R$ and determine therefrom the electron temperature, we obtain the given empirical fit (Fig.~\ref{oiii_temperature.fig}). To reach a perfect correlation, a polynomial correction is applied to the residuals. The fit is valid from 5\,000 to 24\,000\,K.
\begin{equation}
\begin{array}[t]{rcl}
r & = & \log(R)\\
\\
T_{\rm e}\,\, [{\rm K}]& = & 5294 \,(r - 0.848)^{-1} \,\,+ \\
 & &19047 - 7769\,r + 944\,r^2\\
 \end{array}
 \label{te_oiii}
\end{equation}

A widely unnoted determination is the one of Keenan et al. (\cite{Keenan_88}) for [\ion{Ar}{iii}] $\lambda\lambda$\ (7135\AA~+ 7751\AA) / $\lambda$ 5192\AA~from the two lower transitions $^1D_{2}\rightarrow\phantom{l}^3P_{2}$ (1.737$\rightarrow$0.000\,eV),
$^1D_{2}\rightarrow\phantom{l}^3P_{1}$ (1.737$\rightarrow$0.138\,eV) and the upper transition $^1S_{0}\rightarrow\phantom{l}^1D_{2}$ (4.124$\rightarrow$1.737\,eV). One reason for the rare use is certainly the red wavelength of the lower transitions. Older spectrographs did not work well in that domain. A second reason might be the fact that only a plot but no table or formula was given in the original paper. On the other hand, compared to [\ion{O}{iii}], the lower energies of the individual levels make especially the upper transition stronger. Additionally it is not influenced by fluorescence mechanisms.

\begin{equation}
\begin{array}[t]{rcl}
r & = & \log(R)\\
\\
T_{\rm e}\,[{\rm K}] & = & 11960 \,(r - 1.02)^{-1}  \\
 \end{array}
\end{equation}

\begin{figure}
   \centering
   \includegraphics[width=8.8cm]{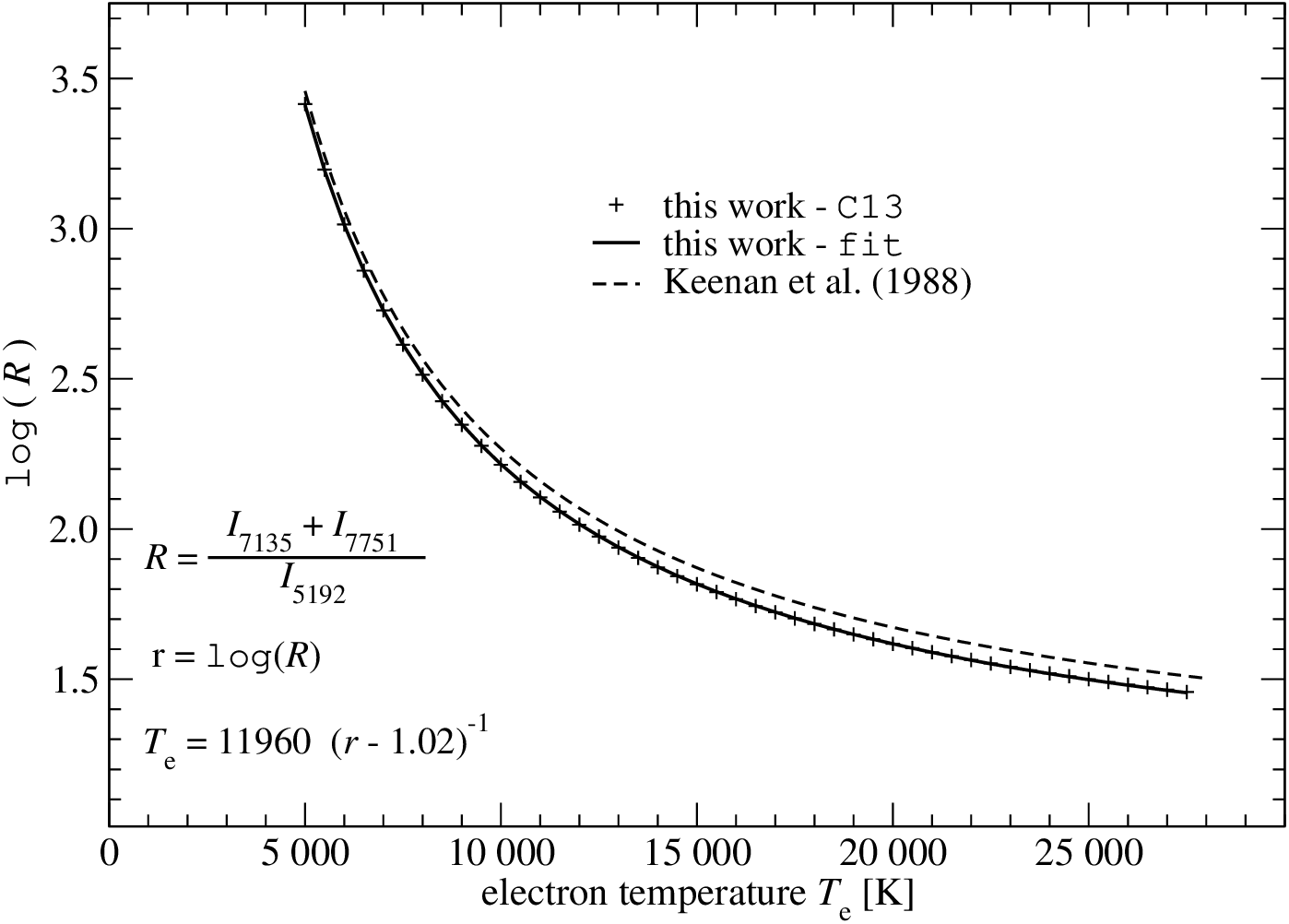}
      \caption{
   The temperature derivation using the  [\ion{Ar}{iii}] lines. The new calibration is about 10\% below the older ones throughout the whole range. The fit overlays the data completely.}
         \label{ariii_temperature.fig}
\end{figure}

\noindent Current atomic data give a temperature about 10\% below that given by the calibration of Keenan et al. (\cite{Keenan_88}) (see Fig.~\ref{ariii_temperature.fig}).

\subsection{The Electron Density}
\begin{figure}
   \centering
   \includegraphics[width=8.8cm]{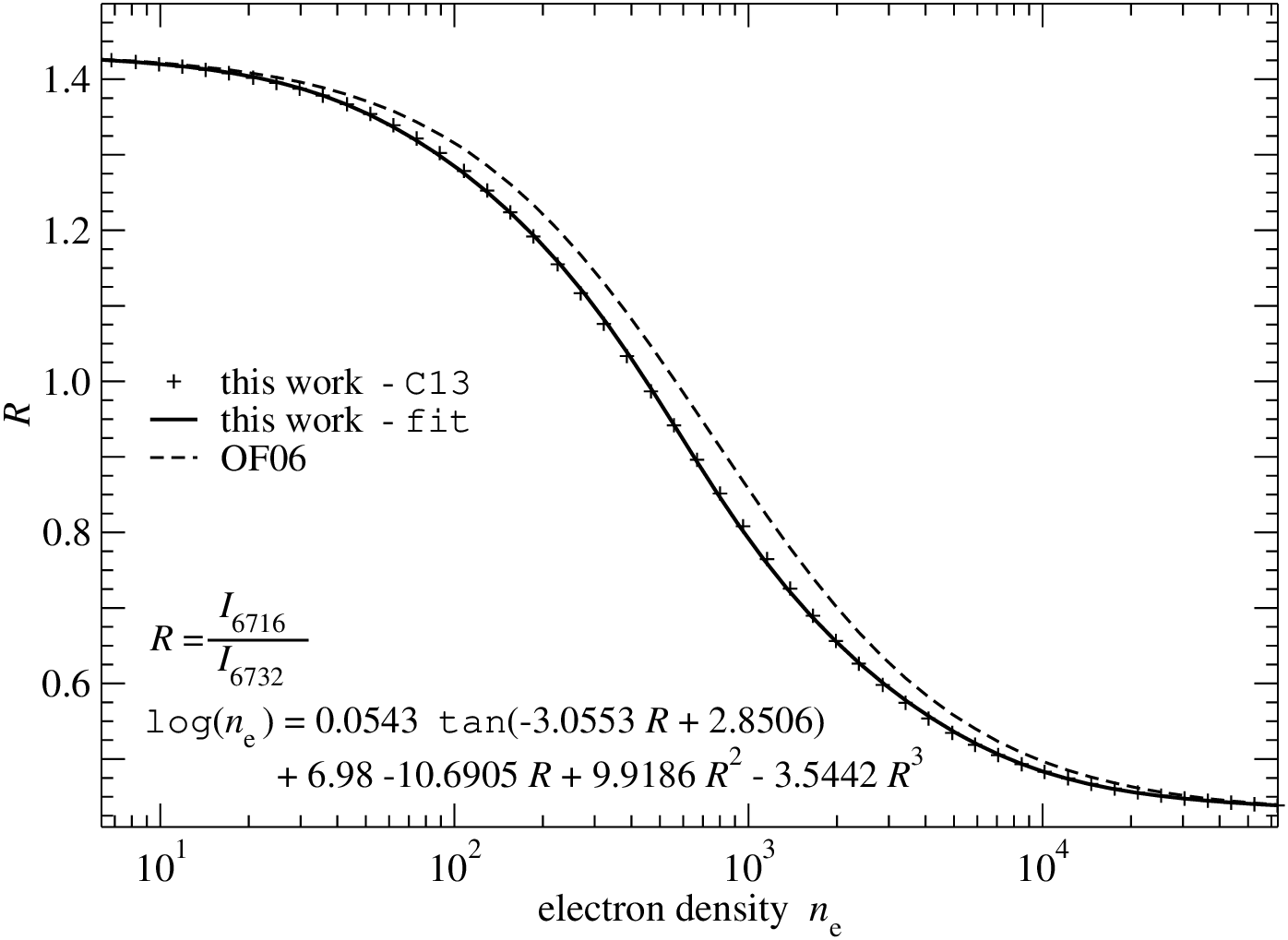}
      \caption{
   The electron density derived by the forbidden lines of [\ion{S}{ii}] at an electron temperature $T_{\rm e} = 10\,000\,{\rm K}$. During the complete well usable range of $40 \le n_{\rm e} \le 10^4$ cm$^{-3}$, the result lies 20 to 22\% lower than the widely used OF06 calibration.}
         \label{sii-density.fig}
   \end{figure}

To our knowledge, the most often used diagnostic diagram for $n_{\rm e}$ is that of [\ion{S}{ii}] $\lambda\lambda$ 6716\AA~/ 6732\AA~with the transitions $^2D_{5/2}^0\rightarrow\phantom{l}^4S_{3/2}^0$ (1.846$\rightarrow$0.000\,eV) and $^2D_{3/2}^0\rightarrow\phantom{l}^4S_{3/2}^0$ (1.842$\rightarrow$0.000\,eV). The calibration published by OF06 is identical to our results in the setup using C08. The new atomic data in C10 and C13 lead to systematically 20--22\% lower electron densities. The curve can be fitted fairly well with an $\arctan$ function. This leads after the inversion (required for the observer) to the $\tan$ function of the resulting function below. To achieve perfect transitions at the lower and upper ends, a polynomial correction was required (Fig.~\ref{sii-density.fig}). Again vice versa and thus a direct tool for the observers, we calculated an empirical fit (see Fig. \ref{sii-density.fig}) for $T_{\rm e} = 10\,000\,{\rm K}$:

\begin{equation}
\begin{array}[t]{rcl}
\log(n_{\rm e}\, [{\rm cm}^{-3}]) & = & 0.0543 \tan(-3.0553\,R + 2.8506)\,\, + \\
 & &6.98 - 10.6905\,R\,\, +  \\
 & &9.9186\,R^2 - 3.5442\,R^3\\
 \end{array}
\end{equation}

\begin{figure}
   \centering
   \includegraphics[width=8.8cm]{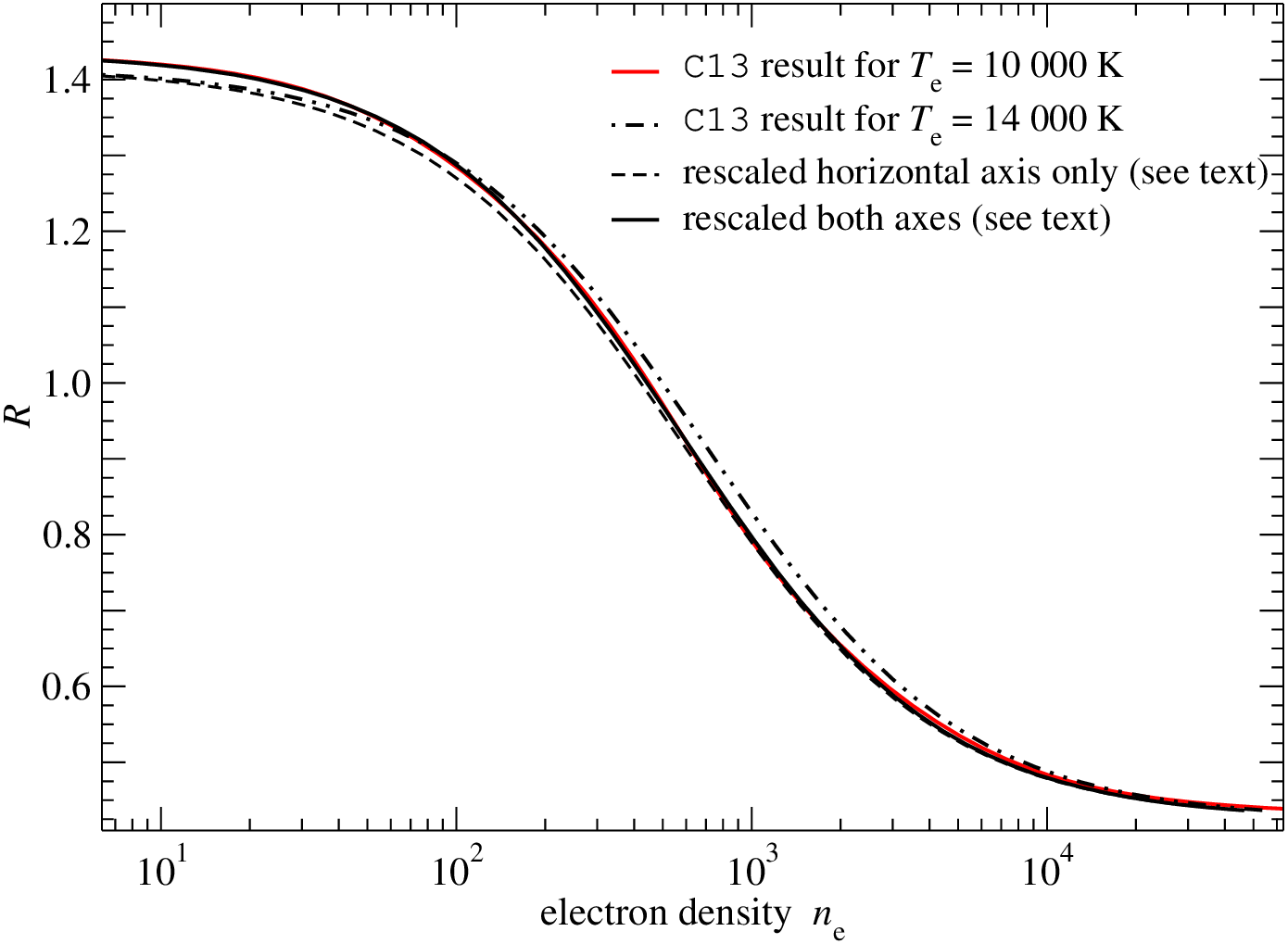}
      \caption{The  scaling of the [\ion{S}{ii}] density for other temperatures. A comparison of the diagram at $T_{\rm e} = 10\,000\,{\rm K}$ (red) and $T_{\rm e} = 14\,000\,{\rm K}$ (dash-dotted line) is shown.  The re-scale of the horizontal axis by $n_{\rm e}(10^4/T_{\rm e})^{1/2}$ (dashed line) is not perfect. An additional scale of the vertical axis results in nearly undistinguishable curves (solid black line). (The color version of the figure is only provided in the electronic version of the journal.)}
         \label{sii-density-scale.fig}
\end{figure}

\noindent We know that in most nebulae we have a co-existence of cold dense clumps, embedded into a hotter surrounding  thin gas. Thus the analysis deriving the pair ($T_{\rm e}$, $n_{\rm e}$) (e.g. for further investigations like abundances) has to take care to obtain information for the same regions within the gas. In most cases these different populations are not spatially resolved in the spectra. As the ionization energy of \ion{S}{i}$\rightarrow$\ion{S}{ii}~(12.20\,eV) is below that of hydrogen, the states can be easily reached by thermal collisional excitation or by absorption of Lyman photons of recombining hydrogen. The next ionization state \ion{S}{ii}$\rightarrow$\ion{S}{iii} requires 23.3\,eV. This can be reached in regions optically thin for UV photons or through photons originating from \ion{He}{} recombination only. So we have to expect, the [\ion{S}{ii}] lines originate in the majority from denser clumps or from the outskirts of nebulae.\protect{\newline}
At other electron temperatures the curves are shifted slightly, but have nearly the same shape. Thus SK89 argued that the effect can be ignored. For higher quality of modern spectra, however, small corrections should be applied. OF06 suggest the re-scale of the horizontal axis by $n_{\rm e}(10^4/T_{\rm e})^{1/2}$. As we tested here (Fig.~\ref{sii-density-scale.fig}), this does not lead to a good result at the low density regime. We applied an additional linear re-scale of the vertical axis by the difference between low density limit $R_\mathrm{low}^{[\ion{S}{ii}]}$ and the temperature independent high density limit $R_\mathrm{high}^{[\ion{S}{ii}]}$. This makes the curves perfectly overlaying.

\noindent A full set of limits for electron temperatures $5\,000 \le T_{\rm e} \le 26\,000\,{\rm K}$ is calculated. It can be described by a quadratic approximation and thus give us the detailed re-scale recipe by:
\begin{equation}
\begin{array}[t]{rcl}
R_\mathrm{low}^{[\ion{S}{ii}]} (T_{\rm e})& = & 1.496 - 0.07442\,\left({T_{\rm e}\,[{\rm K}]\over 10^4}\right) + 0.01225\,\left({T_{\rm e}\,[{\rm K}]\over 10^4}\right)^2\\
\\
R_\mathrm{high}^{[\ion{S}{ii}]} & = & 0.436\\
\\
R_\mathrm{obs}^{[\ion{S}{ii}]} (T_{\rm e}) &\rightarrow&R_\mathrm{re-scaled}^{[\ion{S}{ii}]} \\
\\
\medskip
R_\mathrm{re-scaled}^{[\ion{S}{ii}]} &=&{{R_\mathrm{low}^{[\ion{S}{ii}]} (10 000K)-R_\mathrm{high}^{[\ion{S}{ii}]}}\over
{{R_\mathrm{low}^{[\ion{S}{ii}]} (T_{\rm e})-R_\mathrm{high}^{[\ion{S}{ii}]}}}}\,\,\left({R_\mathrm{obs}^{[\ion{S}{ii}]} (T_{\rm e})-R_\mathrm{high}^{[\ion{S}{ii}]}}\right) + R_\mathrm{high}^{[\ion{S}{ii}]} \\
\\
\medskip
&\rightarrow&n_{\rm e}(10000K)\\
n_{\rm e}(T_{\rm e}) &=& n_{\rm e}(10000K)\,\left({10000\over T_{\rm e}}\right)^{-{1\over2}}

\end{array}
\end{equation}

\begin{figure}
   \centering
   \includegraphics[width=8.8cm]{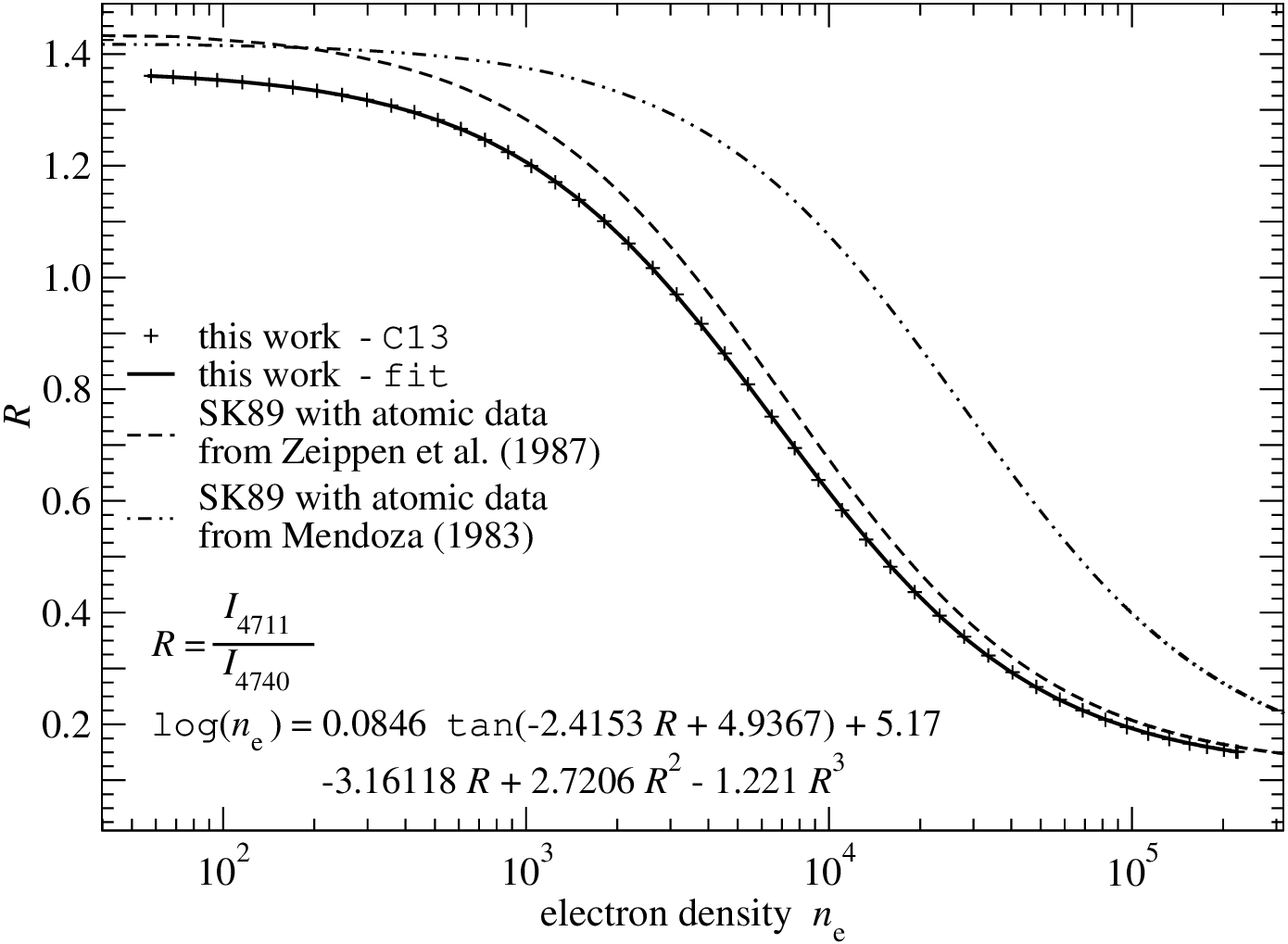}
      \caption{
   The new diagnostic diagram for [\ion{Ar}{iv}]. For comparison the two sets of SK89 are shown. }
         \label{ariv_density.fig}
\end{figure}

\noindent SK89 presented two different calibrations for the electron density using the [\ion{Ar}{iv}] $\lambda\lambda$ 4711\AA~/ 4740\AA~lines with the transitions $^2D_{5/2}^0\rightarrow\phantom{l}^4S_{3/2}^0$ (2.631$\rightarrow$0.000\,eV)  and $^2D_{3/2}^0\rightarrow\phantom{l}^4S_{3/2}^0$ (2.615$\rightarrow$0.000\,eV). They used atomic data sets of Mendoza (\cite{mendoza83}) and Zeippen et al. (\cite{zeippen87}). Our new results are presented in Fig.~\ref{ariv_density.fig}. The differences for the derived $n_{\rm e}$ are fairly high (up to a factor of five). Derived in the same way as above, the fit function is:
\begin{equation}
\begin{array}[t]{rcl}
\log(n_{\rm e} \, [{\rm cm}^{-3}]) & = & 0.0846 \tan(-2.4153\,R + 4.9367)\,\, + \\
 & &5.17 - 3.16118\,R\,\,+\\
 & &2.7206\,R^2 - 1.221\,R^3\\
 \end{array}
\end{equation}

\noindent The ionization energies for \ion{Ar}{i} $\rightarrow$ \ion{Ar}{ii}, \ion{Ar}{ii} $\rightarrow$ \ion{Ar}{iii} and \ion{Ar}{iii} $\rightarrow$ \ion{Ar}{iv} are 15.76\,eV, 27.63\,eV and 40.73\,eV respectively. These levels cannot be reached thermodynamically (equilibrium according to Saha's equation) by the typical electron temperatures in \ion{H}{ii} regions or PNe. Thus they originate from thin hot regions of the plasma, which are optically thin for UV photons. This makes the $n_{\rm e}$([\ion{Ar}{iv}]) more suitable for the combination with a temperature determination by [\ion{O}{iii}] lines, compared to  $n_{\rm e}$([\ion{S}{ii}]).

\noindent For purpose of scaling the low density limits $R_\mathrm{low}^{[\ion{Ar}{iv}]}$ for the electron temperature range $8\,000 \le T_{\rm e} \le 26\,000\,$ K are calculated. At lower temperatures the lines abruptly vanish due to their required high excitation energy. The recipe follows the same way as for [\ion{S}{ii}].
\begin{equation}
\begin{array}[t]{rcl}
R_\mathrm{low}^{[\ion{Ar}{iv}]} (T_{\rm_ e}) & =&  1.4663 - 0.10806\,\left({T_{\rm e}\,[{\rm K}]\over 10^4}\right) + 0.01266\,\left({T_{\rm e}\,[{\rm K}]\over 10^4}\right)^2\\
\\
R_\mathrm{high}^{[\ion{Ar}{iv}]} & = & 0.120\\
\end{array}
\end{equation}

\section{Conclusions}
Using the current state-of-the-art code and modern atomic data, significant differences for the given diagnostic diagrams are found compared to older studies. Thus we encourage the observers to use these new determinations for the interpretation of their data. 
\noindent Moreover we would like to draw the attention of the observers to the use of ions with comparable excitation state for deriving $n_{\rm e}$ and $T_{\rm e}$. Due to the observational constraints [\ion{S}{ii}] or [\ion{O}{ii}] densities are often combined with [\ion{O}{iii}] electron temperatures.
One may draw wrong conclusions by looking at different plasmas. The cold dense clumps, embedded into a hot thin gas, dominate the radiation of the [\ion{S}{ii}] lines. But they are small and thus observationally not resolved. The hot thin gas around the clumps is the origin of the [\ion{O}{iii}] lines. This is well shown in the nonlinear correlations of $T_{\rm e}$ found in data from different ionization potential (Copetti \& Writzl \cite{CoWr02}; Izotov et al. \cite{izotov06}).
Modern spectrographs (e.g. X-Shooter @ ESO) take a very wide range of wavelengths with a single shot at high resolutions. Thus further sets of lines should be investigated to improve accuracy and to widen the range of usable density and temperature parameter space.

\begin{acknowledgements}
      S.{\"O}. is supported by the Austrian \emph{Fonds zur Wissenschaftlichen Forschung, FWF\/} doctoral school project W1227.
\end{acknowledgements}


\begin{thebibliography}{}

\bibitem[2004]{AlGr04}
Aleman, I., \& Gruenwald, R. 2004, ApJ, 607, 865

\bibitem[2002]{CoWr02}
Copetti, M. V. F., \&  Writzl, B. C. 2002, A\&A, 382, 282

\bibitem[1998]{Cloudy_90}
Ferland, G. J., Korista, K. T., Verner, D. A., Ferguson, J. W., Kingdon, J. B., \& Verner, E. M. 1998, PASP, 110, 761

\bibitem[2013]{Cloudy_13}
Ferland, G. J., Porter, R. L., van Hoof, P. A. M., et al. 2013, Rev. Mex. A\&A, 49, 137

\bibitem[2006]{izotov06}
Izotov, Y. I., Stasi{\'n}ska, G., Meynet, G., Guseva, N. G., \& Thuan, T. X. 2006, A\&A, 448, 955

\bibitem[1988]{Keenan_88}
Keenan, F.P., Johnson, C.T., \& Kingston A.E. 1988, A\&A, 202 ,253

\bibitem[2002]{leubner}
Leubner, M. P. 2002, Ap\&SS, 282, 573

\bibitem[2009]{kappa09}
Livadiotis, G., \& McComas, D. J. 2009, Journal of Geophysical Research (Space Physics), 114, 11105

\bibitem[1983]{mendoza83}
Mendoza, C. 1983, IAU Symp., 103, 245

\bibitem[2013]{kappa}
Nicholls, D. C., Dopita, M. A., Sutherland, R. S., Kewley, L. J., \& Palay, E. 2013, ApJS, 207, 21

\bibitem[2012]{nichols12}
Nicholls, D. C., Dopita, M. A., \& Sutherland, R. S. 2012, ApJ, 752, 148

\bibitem[2006]{O_F_06}
Osterbrock, D.~E. \& Ferland, G.~J. 2006, Astrophysics of gaseous Nebulae and active Galactic Nuclei (University Science Books, Sausalito, CA)
	
\bibitem[2012]{palay12}
Palay, E., Nahar, S. N., Pradhan, A. K., \& Eissner, W. 2012, MNRAS, 423, L35

\bibitem[2006]{Pradhan06}
Pradhan, A. K., Montenegro, M. Nahar, S. N., \& Eissner, W. 2006, MNRAS, 366, L6

\bibitem[2013]{BSc}
Proxauf, B. 2013, Upgrading diagnostic diagrams of
forbidden line emission: The influence of modern atomic data, BSc thesis, University Innsbruck, pp. 24

\bibitem[1989]{ArIV}
Stanghellini, L. \&  Kaler, J.B. 1989, ApJ, 343, 811

\bibitem[2000]{vanHoof00}
van Hoof, P. A. M., Van de Steene, G. C., Beintema, D. A., Martin, P. G.; Pottasch, S. R., \& Ferland, G. J. 2000, ApJ, 532, 384

\bibitem[1987]{zeippen87}
Zeippen, C.J., Butler, K., \& Le Bourlot, J. 1987, Astr. Ap., 188, 251

\end{thebibliography}
\end{document}